\documentclass[pra,amsmath,aps,twocolumn,superscriptaddress]{revtex4}
\usepackage{amsmath,mathrsfs,amsbsy,amssymb,lmodern,graphicx,bm,amsthm,amsfonts}
\usepackage{units}
\usepackage{bbm}
\usepackage{multirow,color}
\usepackage{subfigure}
\newcommand{\tr}{{\rm Tr}}
\begin{document}
\title{Triangle inequalities in coherence measures and entanglement concurrence}
\author{Yue Dai}
\affiliation{College of Physics, Optoelectronics and Energy, Soochow University, Suzhou, 215006, China}
\author{Wenlong You}
\affiliation{College of Physics, Optoelectronics and Energy, Soochow University, Suzhou, 215006, China}
\author{Yuli Dong}
\affiliation{College of Physics, Optoelectronics and Energy, Soochow University, Suzhou, 215006, China}
\author{Chengjie Zhang}
\email{zhangchengjie@suda.edu.cn}
\affiliation{College of Physics, Optoelectronics and Energy, Soochow University, Suzhou, 215006, China}
\affiliation{Key Laboratory of Quantum Information, University of Science and Technology of China, CAS, Hefei, 230026, China}

\begin{abstract}
We provide detailed proofs of triangle inequalities in coherence measures and entanglement concurrence. If a rank-$2$ state $\varrho$ can be expressed as a convex combination of two pure states, i.e., $\varrho=p_{1}|\psi_{1}\rangle\langle\psi_{1}|+p_{2}|\psi_{2}\rangle\langle\psi_{2}|$, a triangle inequality can be established as $\big{|}E(|\Psi_{1}\rangle)-E(|\Psi_{2}\rangle)\big{|}\leq E(\varrho)\leq E(|\Psi_{1}\rangle)+E(|\Psi_{2}\rangle)$, where $|\Psi_{1}\rangle=\sqrt{p_{1}}|\psi_{1}\rangle$ and $|\Psi_{2}\rangle=\sqrt{p_{2}}|\psi_{2}\rangle$, $E$ can be considered either coherence measures or entanglement concurrence. This inequality displays mathematical beauty for its similarity to the triangle inequality in plane geometry. An illustrative example is given after the proof.
\end{abstract}
\date{\today}

\maketitle

\section{Introduction}
Coherence and entanglement are two crucial quantum mechanical properties which are widely used in quantum information processing and quantum computation \cite{ci}. While quantum coherence is defined for single systems, quantum entanglement is adopted to describe bipartite or multipartite systems. In earlier research, coherence is usually a main concern of quantum optics. But a new viewpoint is proposed that coherence can be treated as a physical resource, just like entanglement \cite{coh1}. Both coherence and entanglement can be characterized by the resource theory. We review some measures of coherence and entanglement at first.

A widely used measure of coherence is the distance-based measure \cite{coh1}. The starting point for the definition of coherence is the identification of the set $\mathbb{I}$ of incoherent quantum states. The incoherent states are diagonal in the reference basis $\{|i\rangle\}_{i=1}^{d}$ (which are chosen according to the practical physical problem), and take the form
\begin{equation}\label{f0}
\delta=\sum_{i=1}^{d}\delta_{i}|i\rangle\langle i|,
\end{equation}
for a $d$-dimensional Hilbert space. A measure of coherence of a state $\varrho$ can be defined by the minimal distance between $\varrho$ and the set $\mathbb{I}$ of incoherent states, i.e.,
\begin{equation}\label{f01}
C_{D}(\varrho)=\min_{\delta\in\mathbb{I}}D(\varrho,\delta),
\end{equation}
where $D(\varrho, \delta)$ denotes certain distance measures of quantum states. If $\varrho$ is a incoherent state, i.e., $\varrho\in\mathbb{I}$, $C_{D}(\varrho)$ must be zero with $\varrho=\delta$.

Concurrence applies to the measure of entanglement \cite{concurrence1,concurrence2}. The definition of concurrence is based on the convex-roof construction. It is suitable for use in both pure states and mixed states \cite{mixs1,mixs2,mixs3}. Unfortunately, analytical solutions of concurrence can only be obtained in $2$-qubit states ($2\otimes2$ dimensions) \cite{wootter1,wootter2} and some high-dimensional bipartite states with high symmetries, such as isotropic states and Werner states \cite{iso,werner}. For general high-dimensional mixed states, it is not fully explored with only a little knowledge \cite{little1,little2}.

For a general high-dimensional bipartite pure state $\varrho_{_{AB}}=|\psi\rangle\langle\psi|$, which is expanded in a finite $d_{1}\otimes d_{2}$-dimensional Hilbert space $\mathcal{H}_{A}\otimes \mathcal{H}_{B}$, the concurrence is defined as $C(|\psi\rangle)=$\small{$\sqrt{2(1-\tr\varrho\!_{_{A}}^{^{\ \!2}})}$}, with $\varrho\!_{_{A}}\!=\!\tr\!_{_{B}}\varrho\!_{_{A\!B}}$ being the reduced density matrix \cite{cdefine}. Moreover, a pure state can be generally expressed as $|\psi\rangle=\sum_{i=1}^{d_{1}} \sum_{j=1}^{d_{2}} \phi_{ij} |ij\rangle$ ($\phi_{ij}\!\in\!\mathbb{C}$) in the computational bases $|i\rangle$ and $|j\rangle$ of $\mathcal{H}_{A}$ and $\mathcal{H}_{B}$, respectively, where $i=1,...,d_{1}$ and $j=1,...,d_{2}$. Then the squared concurrence of a pure state can be written as \cite{cc}:
\begin{equation}\label{f1}
C^{2}(|\psi\rangle)=\sum_{m=1}^{D_{1}} \sum_{n=1}^{D_{2}}|C_{mn}|^{2}=4\sum_{i<j}^{d_{1}} \sum_{k<l}^{d_{2}}|\phi_{ik}\phi_{jl}-\phi_{il}\phi_{jk}|^{2},
\end{equation}
where $D_{1}=d_{1}(d_{1}-1)/2$, $D_{2}=d_{2}(d_{2}-1)/2$, $C_{mn}=\langle\psi|\widetilde{\psi}_{mn}\rangle$, $|\widetilde{\psi}_{mn}\rangle=(L_{m}\otimes L_{n})|\psi^{*}\rangle$, and $*$ denotes complex conjugate. Here $L_{m}$, $m=1,...,d_{1}(d_{1}-1)/2$ and $L_{n}$, $n=1,...,d_{2}(d_{2}-1)/2$ are the generators of group SO$(d_{1})$ and SO$(d_{2})$: $L_{m}=|i_{m}\rangle\langle j_{m}|-|j_{m}\rangle \langle i_{m}|$ and $L_{n}=|k_{n}\rangle\langle l_{n}|-|l_{n}\rangle \langle k_{n}|$.

The concurrence of a pure state can be easily calculated as zero if the state is separable. For a mixed state $\varrho=\sum_{i} p_{i}|\psi_{i}\rangle\langle\psi_{i}|$, $\sum_{i} p_{i}=1$, the concurrence is defined by the convex-roof \cite{cr1,cr2,cr3,cr4} as follows
\begin{equation}\label{f3}
C(\varrho)\equiv \min_{\{p_{i},|\psi_{i}\rangle\}}\sum_{i} p_{i}C(|\psi_{i}\rangle).
\end{equation}
The minimization is taken over all possible decompositions of $\varrho$ into pure states.
For a $2$-qubit mixed state $\varrho$, an analytic solution of concurrence can be calculated:
\begin{equation}\label{f33}
C(\varrho)=\max \{0,\lambda_{1}-\lambda_{2}-\lambda_{3}-\lambda_{4}\},
\end{equation}
where $\{\lambda_{i}\}$ are the eigenvalues, in decreasing order, of the Hermitian matrix $R\equiv\sqrt{\sqrt{\varrho}\widetilde{\varrho}\sqrt{\varrho}}$, $\widetilde{\varrho}$ is the spin-flipped state $\widetilde{\varrho}=(\sigma_{y}\otimes\sigma_{y})\varrho^{*}(\sigma_{y}\otimes\sigma_{y})$. It is worth noticing that $\{\lambda_{i}\}$ are also the singular values of a complex symmetric matrix $\tau$, where $\tau_{ij}=\langle\upsilon_{i}|\widetilde{\upsilon}_{j}\rangle$. The states $|\upsilon_{i}\rangle$ are the eigenstates of $\varrho$ \cite{wootter2}.
For a high-dimensional mixed state, the minimum decomposition is very cumbersome to detect. We usually provide a bound for concurrence to analyze mixed states. Until now, several bounds of concurrence have been introduced \cite{bound1,bound2,bound3,bound4,bound5,bound6,bound7,bound8,bound9,bound10,bound11,bound12,bound13,bound14,bound15,bound16,bound17,bound18,bound19,bound20,bound21,bound22,bound23,bound24,bound25,bound26}.

In our research, we find that if a rank-$2$ mixed state $\varrho$ can be expressed as a convex combination of two pure states, i.e., $\varrho=p_{1}|\psi_{1}\rangle\langle\psi_{1}|+p_{2}|\psi_{2}\rangle\langle\psi_{2}|$, ($|\psi_{1}\rangle$ and $|\psi_{2}\rangle$ are linearly independent), the triangle inequality can be established as
\begin{equation}\label{f4}
\big{|}E(|\Psi_{1}\rangle)-E(|\Psi_{2}\rangle)\big{|}\leq  E(\varrho)\leq E(|\Psi_{1}\rangle)+E(|\Psi_{2}\rangle),
\end{equation}
where $|\Psi_{1}\rangle=\sqrt{p_{1}}|\psi_{1}\rangle$ and $|\Psi_{2}\rangle=\sqrt{p_{2}}|\psi_{2}\rangle$, $E$ can be considered either coherence measures or entanglement concurrence. It is similar to the triangle inequality in geometry. In Sec. \uppercase\expandafter{\romannumeral2} and \uppercase\expandafter{\romannumeral3}, the triangle inequality is proven in coherence measures and entanglement concurrence, respectively.

\section{Triangle inequalities in coherence measures}
Several measures of coherence are proposed based on the distance-based measures, such as the relative entropy, the $l_{1}$ norm and the trace norm \cite{coh1}. Here we only focus on the form of the $l_{1}$ norm
\begin{equation}\label{f02}
C_{l_{1}}(\varrho)=\min_{\delta\in\mathbb{I}}\|\varrho-\delta\|_{l_{1}}=\sum_{i\neq j} \big{|}\langle i|\varrho|j\rangle\big{|},
\end{equation}
which is equal to sum of the absolute values of all off-diagonal elements of $\varrho$.

For mixed states, the convex-roof $l_{1}$ norm is adopted as a different measure of coherence \cite{cohcon}. The convex-roof construction is used to define concurrence and some other measures of entanglement previously. Taking into account the resource theory for coherence, we can make use of the convex-roof construction to measure coherence similar to its application in entanglement. The convex-roof $l_{1}$ norm of a mixed state $\varrho$ takes the form
\begin{equation}\label{f03}
\widetilde{C}_{l_{1}}(\varrho)\equiv \min_{\{p_{i},|\psi_{i}\rangle\}}\sum_{i} p_{i}C_{l_{1}}(|\psi_{i}\rangle),
\end{equation}
where the minimization is taken over all pure state decompositions of $\varrho=\sum_{i} p_{i}|\psi_{i}\rangle\langle\psi_{i}|$, $\sum_{i} p_{i}=1$. $C_{l_{1}}(|\psi_{i}\rangle)$ is the $l_{1}$ norm of the state $|\psi_{i}\rangle$. It is very similar to the entanglement concurrence.

We begin with a general triangle inequality in the $l_{1}$ norm measure of coherence.

\textit{Theorem 1.} If a state $\varrho$ can be expressed as a convex combination of two states $\varrho=p_{1}\varrho_{1}+p_{2}\varrho_{2}$,  the $l_{1}$ norm of $\varrho$, i.e. $C_{l_{1}}(\varrho)$, satisfies the triangle inequality
\begin{equation}\label{f444}
\big{|}C_{l_{1}}(p_{1}\varrho_{1})-C_{l_{1}}(p_{2}\varrho_{2})\big{|}\leq C_{l_{1}}(\varrho)\leq C_{l_{1}}(p_{1}\varrho_{1})+C_{l_{1}}(p_{2}\varrho_{2}).
\end{equation}

\textit{Proof.} The $l_{1}$ norm of $\varrho$ can be expressed as
\begin{align}\label{f45}
C_{l_{1}}(\varrho)&=C_{l_{1}}(p_{1}\varrho_{1}+p_{2}\varrho_{2}) \nonumber\\
&=\sum_{i\neq j}\big{|}p_{1}\langle i|\varrho_{1}|j\rangle+p_{2}\langle i|\varrho_{2}|j\rangle\big{|}.
\end{align}
Considering absolute value inequality, the right hand side of Eq. (\ref{f45}) should conform to the inequality
\begin{small}
\begin{align}\label{f46}
&\bigg{|}\sum_{i\neq j}p_{1}\big{|}\langle i|\varrho_{1}|j\rangle\big{|}-\sum_{i\neq j}p_{2}\big{|}\langle i|\varrho_{2}|j\rangle\big{|}\bigg{|}\nonumber\\
&\leq\sum_{i\neq j}\big{|}p_{1}\langle i|\varrho_{1}|j\rangle+p_{2}\langle i|\varrho_{2}|j\rangle\big{|} \nonumber\\
&\leq\sum_{i\neq j}p_{1}\big{|}\langle i|\varrho_{1}|j\rangle\big{|}+\sum_{i\neq j}p_{2}\big{|}\langle i|\varrho_{2}|j\rangle\big{|}.
\end{align}
\end{small}
According to the definition of the $l_{1}$ norm of coherence, $C_{l_{1}}(p_1\varrho_{1})=\sum_{i\neq j}p_{1}\big{|}\langle i|\varrho_{1}|j\rangle\big{|}$ and $C_{l_{1}}(p_2\varrho_{2})=\sum_{i\neq j}p_{2}\big{|}\langle i|\varrho_{2}|j\rangle\big{|}$. Then the triangle inequality can be established
\begin{equation}\label{f47}
\big{|}C_{l_{1}}(p_{1}\varrho_{1})-C_{l_{1}}(p_{2}\varrho_{2})\big{|}\leq C_{l_{1}}(\varrho)\leq C_{l_{1}}(p_{1}\varrho_{1})+C_{l_{1}}(p_{2}\varrho_{2}).
\end{equation}
Note that for arbitrary $\varrho$, $\varrho_{1}$ and $\varrho_{2}$ can be alternatively pure states or mixed states.

If $\varrho$ is a rank-$2$ mixed states and the decomposition parts of $\varrho$ are two pure states $|\psi_{1}\rangle$ and $|\psi_{2}\rangle$, then the convex-roof $l_{1}$ norm of $\varrho$ should also satisfy the triangle inequality.

\textit{Theorem 2.} For a rank-$2$ mixed state $\varrho$, if it can be decomposed into two pure states $|\psi_{1}\rangle$ and $|\psi_{2}\rangle$ with linear independence: $\varrho=p_{1}|\psi_{1}\rangle\langle\psi_{1}|+p_{2}|\psi_{2}\rangle\langle\psi_{2}|$, let $|\Psi_{1}\rangle=\sqrt{p_{1}}|\psi_{1}\rangle$ and $|\Psi_{2}\rangle=\sqrt{p_{2}}|\psi_{2}\rangle$, the convex-roof $l_{1}$ norm of $\varrho$, i.e. $\widetilde{C}_{l_{1}}(\varrho)$ satisfies the triangle inequality:
\begin{equation}\label{f48}
\big{|}\widetilde{C}_{l_{1}}(|\Psi_{1}\rangle)-\widetilde{C}_{l_{1}}(|\Psi_{2}\rangle)\big{|}\leq \widetilde{C}_{l_{1}}(\varrho)\leq \widetilde{C}_{l_{1}}(|\Psi_{1}\rangle)+\widetilde{C}_{l_{1}}(|\Psi_{2}\rangle).
\end{equation}
where $\widetilde{C}_{l_{1}}(|\Psi_{1}\rangle)$ and $\widetilde{C}_{l_{1}}(|\Psi_{2}\rangle)$ are the convex-roof $l_{1}$ norm of $|\Psi_{1}\rangle$ and $|\Psi_{2}\rangle$ respectively.

\textit{Proof.} Note that the convex-roof $l_{1}$ norm equals $l_{1}$ norm of coherence for pure states. So Eq. (\ref{f48}) can be read as
 \begin{equation}\label{f49}
\big{|}C_{l_{1}}(|\Psi_{1}\rangle)-C_{l_{1}}(|\Psi_{2}\rangle)\big{|}\leq \widetilde{C}_{l_{1}}(\varrho)\leq C_{l_{1}}(|\Psi_{1}\rangle)+C_{l_{1}}(|\Psi_{2}\rangle).
\end{equation}
 The right hand side of Eq. (\ref{f49}) can be proven by the definition of the convex-roof $l_{1}$ norm. $\widetilde{C}_{l_{1}}(\varrho)$ is a sum of the minimal decomposition of $\varrho$, $C_{l_{1}}(|\Psi_{1}\rangle)+C_{l_{1}}(|\Psi_{2}\rangle)$ can be regarded as a sum of a general decomposition of $\varrho$. So the right hand side is established.

The convex-roof $l_{1}$ norm is not less than the $l_{1}$ norm of coherence for mixed state, i.e., $C_{l_{1}}(\varrho) \leq \widetilde{C}_{l_{1}}(\varrho)$ for any state $\varrho$ \cite{cohcon}. Here we give a simple proof of this corollary.

Assume that $\varrho=\sum_{k}q_{k}|\phi_{k}\rangle\langle\phi_{k}|$, $\sum_{k}q_{k}|\phi_{k}\rangle\langle\phi_{k}|$ is the sum of minimal decomposition, the convex-roof $l_{1}$ norm follows
\begin{align}\label{f50}
\widetilde{C}_{l_{1}}(\varrho)=\sum_{k}q_{k}C_{l_{1}}(|\phi_{k}\rangle)&=\sum_{k}q_{k}\sum_{i\neq j}\big{|}\langle i|\phi_{k}\rangle\langle\phi_{k}|j\rangle\big{|}\nonumber\\
&\geq \sum_{i\neq j}\big{|}\langle i|\sum_{k} q_{k}|\phi_{k}\rangle\langle\phi_{k}|j\rangle\big{|}\nonumber\\
&=\sum_{i\neq j}\big{|}\langle i|\varrho|j\rangle\big{|}\nonumber\\
&=C_{l_{1}}(\varrho)
\end{align}

Considering theorem 1, it is correct that $\big{|}C_{l_{1}}(|\Psi_{1}\rangle)-C_{l_{1}}(|\Psi_{2}\rangle)\big{|}\leq C_{l_{1}}(\varrho) \leq \widetilde{C}_{l_{1}}(\varrho)$. The proof is over.

\textit{Remark.} The convex-roof $l_{1}$ norm was named coherence concurrence for its similarity to entanglement concurrence \cite{cohcon}. It is interesting that both coherence concurrence and entanglement concurrence satisfy the triangle inequality with a rank-$2$ state. There is a potential question that whether other measures of coherence, such as the relative entropy and the trace norm, are suitable to the triangle inequality. The triangle inequality in the $l_{1}$ norm is simple to prove for the easy computation of the $l_{1}$ norm, but other measures are not so simple to be analyzed.

\section{Triangle inequality in entanglement concurrence}
As we mentioned in the introduction section that the matrix $\tau$ is a complex symmetric matrix, and $\{\lambda_{i}\}$ can be alternatively considered as the singular values of $\tau$. At the beginning, we propose a lemma for complex symmetric matrices which is helpful to prove the triangle inequality.

\subsection{Two-qubit states}

\textit{Lemma 1.} For a $2$ by $2$ complex symmetric matrix with nonzero diagonal elements $x_{1}$, $x_{2}$ and singular values $\sigma_{1}$, $\sigma_{2}$ (set $\sigma_{1}\geq \sigma_{2}$), the inequality can be established as
\begin{equation}\label{f5}
\big{|}|x_{1}|-|x_{2}|\big{|}\leq\sigma_{1}-\sigma_{2}.
\end{equation}

\textit{Proof.} A $2$ by $2$ complex symmetric matrix $\tau$ can be expressed as $\tau=U\Sigma U^{T}$ by Singular Value Decomposition (SVD), $UU^{\dag}=I$ and $\Sigma=\mathrm{diag}\{\sigma_{1},\sigma_{2}\}$. The $2$ by $2$ unitary matrix $U$ can read
\begin{equation}\label{f6}
U=e^{ir}\left( \begin{matrix}
	a&		b\\
	-b^*&		a^*\\
\end{matrix} \right),
\end{equation}
where $|a|^{2}+|b|^{2}=1, r^{*}=r$. Then the matrix $\tau$ reads
\begin{equation}\label{f7}
\tau=e^{i2r}\left( \begin{matrix}
	a&		b\\
	-b^*&		a^*\\
\end{matrix} \right) \left( \begin{matrix}
	\sigma _1&		0\\
	0&		\sigma _2\\
\end{matrix} \right) \left( \begin{matrix}
	a&		-b^*\\
	b&		a^*\\
\end{matrix} \right).
\end{equation}

Based on Eq. (\ref{f7}), the diagonal elements $x_{1}$ and $x_{2}$ can be expressed as $x_{1}=e^{i2r}(\sigma_{1}a^{2}+\sigma_{2}b^{2})$ and $x_{2}=e^{i2r}(\sigma_{1}b^{* ^{2}}+\sigma_{2}a^{* ^{2}})$. For $|a|^{2}+|b|^{2}=1$, set $a=e^{i\theta_{1}}\cos\alpha$ and $b=e^{i\theta_{2}}\sin\alpha$. Then we have
\begin{widetext}
\begin{align}\label{f8}
\big{|}|x_{1}|-|x_{2}|\big{|}=&\Big{|}\sqrt{(\sigma_{1}e^{i2\theta_{1}}\cos^{2}\alpha+\sigma_{2}e^{i2\theta_{2}}\sin^{2}\alpha)(\sigma_{1}e^{-i2\theta_{1}}\cos^{2}\alpha+\sigma_{2}e^{-i2\theta_{2}}\sin^{2}\alpha)}\nonumber\\
&-\sqrt{\sigma_{1}e^{-i2\theta_{2}}\sin^{2}\alpha+\sigma_{2}e^{-i2\theta_{1}}\cos^{2}\alpha)(\sigma_{1}e^{i2\theta_{2}}\sin^{2}\alpha+\sigma_{2}e^{i2\theta_{1}}\cos^{2}\alpha}\Big{|}\nonumber\\
\leq&\frac{\big{|}(\sigma_{1}^{2}-\sigma_{2}^{2})(\cos^{2}\alpha-\sin^{2}\alpha)\big{|}}{\sqrt{(\sigma_{1}\cos^{2}\alpha-\sigma_{2}\sin^{2}\alpha)^{2}}+\sqrt{(\sigma_{1}\sin^{2}\alpha-\sigma_{2}\cos^{2}\alpha)^{2}}}.\nonumber\\
\end{align}
\end{widetext}

For the right hand side of Eq. (\ref{f8}), we discuss the results with four situations:

\romannumeral1. If $\cos^{2}\alpha\geq\sin^{2}\alpha$ and $\sigma_{1}\sin^{2}\alpha\geq\sigma_{2}\cos^{2}\alpha$:
\begin{equation}\label{f9}
R.H.S=(\sigma_{1}+\sigma_{2})(\cos^{2}\alpha-\sin^{2}\alpha)\leq\sigma_{1}-\sigma_{2}.
\end{equation}

\romannumeral2. If $\cos^{2}\alpha\geq\sin^{2}\alpha$ and $\sigma_{1}\sin^{2}\alpha\leq\sigma_{2}\cos^{2}\alpha$:
\begin{equation}\label{f10}
R.H.S=\sigma_{1}-\sigma_{2}.
\end{equation}

\romannumeral3. If $\sin^{2}\alpha\geq\cos^{2}\alpha$ and $\sigma_{1}\cos^{2}\alpha\geq\sigma_{2}\sin^{2}\alpha$:
\begin{equation}\label{f11}
R.H.S=(\sigma_{1}+\sigma_{2})(\sin^{2}\alpha-\cos^{2}\alpha)\leq\sigma_{1}-\sigma_{2}.
\end{equation}

\romannumeral4. If $\sin^{2}\alpha\geq\cos^{2}\alpha$ and $\sigma_{1}\cos^{2}\alpha\leq\sigma_{2}\sin^{2}\alpha$:
\begin{equation}\label{f12}
R.H.S=\sigma_{1}-\sigma_{2}.
\end{equation}

In conclusion, for a $2$ by $2$ complex symmetric matrix, the inequality $\big{|}|x_{1}|-|x_{2}|\big{|}\leq\sigma_{1}-\sigma_{2}$ is always correct.

\textit{Remark.} Lemma 1 is the crux of this paper. It can be derived by Thompson theorem \cite{Tho}. But it is so important that we give a detailed proof by ourself. This lemma can be directly applied in a rank-$2$ mixed state $\varrho=p_{1}|\psi_{1}\rangle\langle\psi_{1}|+p_{2}|\psi_{2}\rangle\langle\psi_{2}|$. For the matrix $\tau$, the diagonal elements ($d_{1}$ and $d_{2}$) are the concurrence of $\sqrt{p_{1}}|\psi_{1}\rangle$ and $\sqrt{p_{2}}|\psi_{2}\rangle$. The difference of the two singular values ($\sigma_{1}-\sigma_{2}$) of $\tau$ is the concurrence of the mixed state $\varrho$.

\textit{Theorem 3.} For a $2$-qubit mixed state $\varrho$, if it can be decomposed into two pure states $|\psi_{1}\rangle$ and $|\psi_{2}\rangle$ with linear independence: $\varrho=p_{1}|\psi_{1}\rangle\langle\psi_{1}|+p_{2}|\psi_{2}\rangle\langle\psi_{2}|$, let $|\Psi_{1}\rangle=\sqrt{p_{1}}|\psi_{1}\rangle$ and $|\Psi_{2}\rangle=\sqrt{p_{2}}|\psi_{2}\rangle$, the bound of concurrence $C(\varrho)$ satisfies the triangle inequality:
\begin{equation}\label{f13}
\big{|}C(|\Psi_{1}\rangle)-C(|\Psi_{2}\rangle)\big{|}\leq C(\varrho)\leq C(|\Psi_{1}\rangle)+C(|\Psi_{2}\rangle).
\end{equation}
where $C(|\Psi_{1}\rangle)$ and $C(|\Psi_{2}\rangle)$ are the concurrences of $|\Psi_{1}\rangle$ and $|\Psi_{2}\rangle$ respectively.

\textit{Proof.} The proof of the right hand side follows \textit{Theorem 2}. We give a detailed proof of the left side of the formula below.

For a mixed state $\varrho=|\Psi_{1}\rangle\langle\Psi_{1}|+|\Psi_{2}\rangle\langle\Psi_{2}|$, the complex symmetric matrix $\tau$ is established as
\begin{equation}\label{f14}
\tau =\left( \begin{matrix}
	\langle \Psi _1 | \widetilde{\Psi }_1 \rangle&		\langle \Psi _1 | \widetilde{\Psi }_2 \rangle\\
	\langle \Psi _2 | \widetilde{\Psi }_1 \rangle&		\langle \Psi _2 | \widetilde{\Psi }_2 \rangle\\
\end{matrix} \right) ,
\end{equation}
where $C(|\Psi_{1}\rangle)=|\langle \Psi _1 | \widetilde{\Psi }_1 \rangle|$ and $C(|\Psi_{2}\rangle)=|\langle \Psi _2 | \widetilde{\Psi }_2 \rangle|$. By the lemma 1, the inequality is established:
\begin{align}\label{f15}
\big{|}C(|\Psi_{1}\rangle)-C(|\Psi_{2}\rangle)\big{|}&=\big{|}|\langle \Psi _1 | \widetilde{\Psi }_1 \rangle|-|\langle \Psi _2 | \widetilde{\Psi }_2 \rangle|\big{|} \nonumber\\
&\leq\lambda_{1}-\lambda_{2} \nonumber\\
&=C(\varrho),
\end{align}
where $\lambda_{1}$ and $\lambda_{2}$ are the singular values of $\tau$, $\lambda_{1}\geq\lambda_{2}$. Theorem 3 is finished.

\textit{Remark.} Consider a rank-$S$ state $\varrho=\sum_{i=1}^{K}p_{i}|\psi_{i}\rangle\langle \psi_{i}|$, where the number $K$ is called the cardinality of the ensemble. Necessarily, $K$ cannot be smaller than the rank $S$. In Theorem 3, the rank of the density matrix $\varrho$ in Eq. (\ref{f13}) equals $2$ and the rank of matrix $R\equiv \sqrt{\sqrt{\varrho}\widetilde{\varrho}\sqrt{\varrho}}$ equals $2$. Eq. (\ref{f33}) is rewritten to suit the rank-$2$ state:$C(\varrho)=\lambda_{1}-\lambda_{2}, \lambda_{1}\geq\lambda_{2}$ \cite{wootter2}.

Theorem 3 can be used to detect entanglement of a $2$-qubit mixed state. If the concurrences of $|\Psi_{1}\rangle$ and $|\Psi_{2}\rangle$ are different, entanglement must exist in the mixed state. We extend the triangle inequality from $2\otimes 2$ dimensions to $d_{1}\otimes d_{2}$ dimensions in theorem 4.

\subsection{Bipartite high-dimensional states}

\textit{Theorem 4.} For a $d_{1}\otimes d_{2}$-dimensional mixed state $\varrho$, if it can be decomposed into two pure states $|\psi_{1}\rangle$ and $|\psi_{2}\rangle$ which are linearly independent, the triangle inequality Eq. (\ref{f13}) still holds.

\textit{Proof.} For a $d_{1}\otimes d_{2}$-dimensional mixed state $\varrho$, the concurrence $C(\varrho)$ satisfies \cite{bound6}:
\begin{equation}\label{f16}
\sum_{m=1}^{d_{1}(d_{1}-1)/2} \sum_{n=1}^{d_{2}(d_{2}-1)/2} C_{mn}^{2}\leq C^{2}(\varrho),
\end{equation}
where $C_{mn}=\lambda_{1}^{mn}-\lambda_{2}^{mn}$ (set $\lambda_{1}^{mn}\geq\lambda_{2}^{mn})$, $\lambda_{1}^{mn}$ and $\lambda_{2}^{mn}$ are the singular values of the matrix $\tau_{mn}$:
\begin{equation}\label{f17}
\tau_{mn} =\left( \begin{matrix}
	\langle \Psi _1 |L_{m}\otimes L_{n}| \Psi _1^{*} \rangle&		\langle \Psi _1 |L_{m}\otimes L_{n}| \Psi _2^{*} \rangle\\
	\langle \Psi _2 |L_{m}\otimes L_{n}| \Psi _1^{*} \rangle&		\langle \Psi _2 |L_{m}\otimes L_{n}| \Psi _2^{*} \rangle\\
\end{matrix} \right) .
\end{equation}
Eq. (\ref{f16}) provides a lower bound of squared concurrence of $\varrho$ \cite{bound6}.

By the lemma 1, the diagonal elements and the singular values of the complex symmetric matrix $\tau_{mn}$ satisfy
\begin{equation}\label{f18}
\big{|}|\langle \Psi _1 |L_{m}\otimes L_{n}| \Psi _1^{*} \rangle|-|\langle \Psi _2 |L_{m}\otimes L_{n}| \Psi _2^{*} \rangle|\big{|}\leq\lambda_{1}^{mn}-\lambda_{2}^{mn}.
\end{equation}
The sum of all squared $m,n$ items of Eq.(\ref{f18}) is given as
\begin{equation}\label{f19}
\sum_{m,n} \big{|}|\langle \Psi _1 |L_{m}\otimes L_{n}| \Psi _1^{*} \rangle|-|\langle \Psi _2 |L_{m}\otimes L_{n}| \Psi _2^{*} \rangle|\big{|}^{2}\leq C^{2}(\varrho).
\end{equation}

For simplicity, we set $\langle \Psi _1 |L_{m}\otimes L_{n}| \Psi _1^{*} \rangle=T_{11}^{mn}$, $\langle \Psi _2 |L_{m}\otimes L_{n}| \Psi _2^{*} \rangle=T_{22}^{mn}$ and $\sum_{m,n}$ instead of $\sum_{m=1}^{1,...,d_{1}(d_{1}-1)/2} \sum_{n=1}^{1,...,d_{2}(d_{2}-1)/2}$. Then the left hand side of Eq. (\ref{f19}) is calculated:
\begin{align}\label{f20}
L.H.S&=\sum_{m,n}(|T_{11}^{mn}|^{2}+|T_{22}^{mn}|^{2}-2|T_{11}^{mn}||T_{22}^{mn}|)\nonumber\\
&\geq\sum_{m,n}|T_{11}^{mn}|^{2}+\sum_{m,n}|T_{22}^{mn}|^{2}\nonumber\\
&-2\sqrt{(\sum_{m,n}|T_{11}^{mn}|^{2})(\sum_{m,n}|T_{22}^{mn}|^{2})}.
\end{align}
where the inequality $\sum_{i}a_{i}b_{i}\leq\sqrt{\sum_{i}a_{i}^{2}\sum_{i}b_{i}^{2}}$ is applied.

It is worth noticing that $|\Psi _1\rangle$ and $|\Psi _2\rangle$ are pure states so that $\sum_{m,n}|T_{11}^{mn}|^{2}=C^2(|\Psi _1\rangle)$ and $\sum_{m,n}|T_{22}^{mn}|^{2}=C^2(|\Psi _2\rangle)$. The right hand side of Eq.(\ref{f20}) reads
\begin{align}\label{f21}
R.H.S&=C^{2}(|\Psi _1\rangle)+C^{2}(|\Psi _2\rangle)-2C(|\Psi _1\rangle)C(|\Psi _2\rangle)\nonumber\\
&=\big{|}C(|\Psi _1\rangle)-C(|\Psi _2\rangle)\big{|}^{2}.
\end{align}
So the inequality $|C(|\Psi _1\rangle)-C(|\Psi _2\rangle)|\leq C(\varrho)$ is tenable.

\textit{Remark.} The triangle inequality in concurrence reveals the relation between pure states and mixed states. Bipartite mixed states are inclined to entanglement. If a bipartite mixed state can decomposed into two pure states and the concurrences of the two pure states are different, this mixed state must have entanglement.

\begin{figure}[htb]
\includegraphics[scale=0.31]{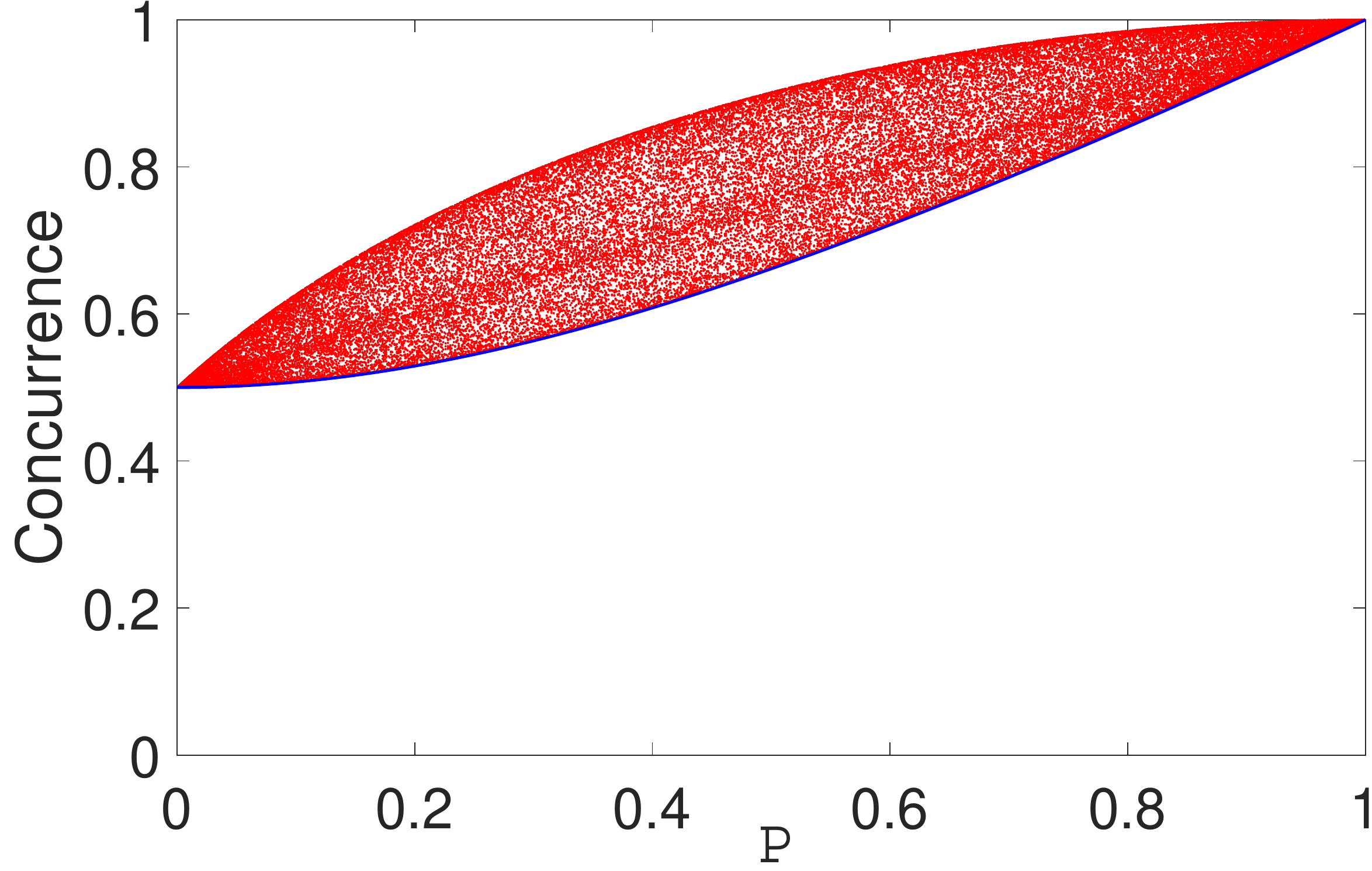}
\caption{The blue line indicates $C(\varrho)$. Red points represent $C(|\Psi_{1}\rangle)+C(|\Psi_{2}\rangle)$ with each point indicating one kind of pure state decomposition for $\varrho$. All red points are distributed above the blue line with a same $P$, implying $C(\varrho)\leq C(|\Psi_{1}\rangle)+C(|\Psi_{2}\rangle)$.}\label{fig1}
\end{figure}

\subsection{Example}

\begin{figure}[htb]
\includegraphics[scale=0.35]{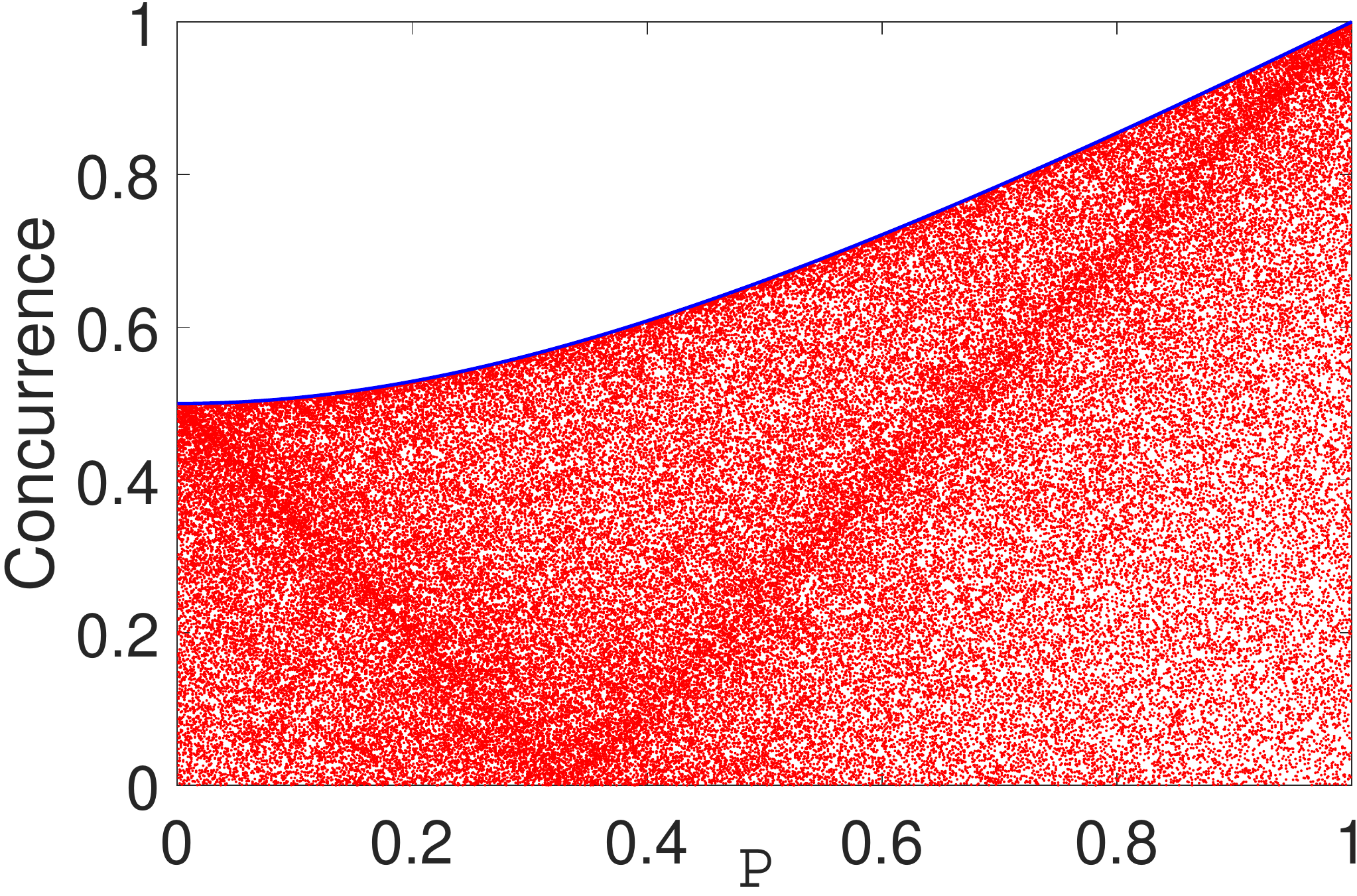}
\caption{The blue line indicates $C(\varrho)$. Red points represent $\big{|}C(|\Psi_{1}\rangle)-C(|\Psi_{2}\rangle)\big{|}$ with  each point indicating one kind of pure state decomposition for $\varrho$. All points points are distributed below the blue line with a same $P$, implying $\big{|}C(|\Psi_{1}\rangle)-C(|\Psi_{2}\rangle)\big{|}\leq C(\varrho)$.}\label{fig2}
\end{figure}

Consider a $2$-qubit mixed state $\varrho=P|\psi_{1}\rangle \langle\psi_{1}|+(1\!-\!P)|\psi_{2}\rangle \langle\psi_{2}|$, where
\begin{align}\label{f22}
&|\psi_{1}\rangle=\sqrt{\frac{3}{8}}(|00\rangle+|11\rangle)+\sqrt{\frac{1}{8}}i(|01\rangle+|10\rangle), \nonumber\\
&|\psi_{2}\rangle=\sqrt{\frac{3}{8}}(|00\rangle+|11\rangle)+\sqrt{\frac{1}{8}}(|01\rangle+|10\rangle)
\end{align}
$|\psi_{1}\rangle$ and $|\psi_{2}\rangle$ are pure states and each qubit has two orthonormal bases $|0\rangle$ and $|1\rangle$. Let $|\Psi_{1}\rangle=\sqrt{P}|\psi_{1}\rangle$, $|\Psi_{2}\rangle=\sqrt{1\!-\!P}|\psi_{2}\rangle$. $C(\varrho)$, $C(|\Psi_{1}\rangle)$ and $C(|\Psi_{2}\rangle)$ are computed with $P$ going from $0$ to $1$. $C(\varrho)$ and $C(|\Psi_{1}\rangle)+C(|\Psi_{2}\rangle)$ are plotted on the vertical y-axis against $P$ on the horizontal x-axis in Fig. \ref{fig1}. $C(\varrho)$ and $\big{|}C(|\Psi_{1}\rangle)-C(|\Psi_{2}\rangle)\big{|}$ are plotted in Fig. \ref{fig2}.

In Fig. \ref{fig1}, the blue line indicates $C(\varrho)$. Red points in the diagram represent $C(|\Psi_{1}\rangle)+C(|\Psi_{2}\rangle)$ with each point indicating one kind of pure state decomposition for $\varrho$. New decomposition, for instance $\varrho=|\Psi'_1\rangle\langle\Psi'_1|+|\Psi'_2\rangle\langle\Psi'_2|$, related to the decomposition $\{|\Psi_1\rangle,|\Psi_2\rangle\}$ by a random unitary transformation:
\begin{equation}\label{f222}
U=\left( \begin{matrix}
	\cos\theta e^{i\gamma}&		\sin\theta e^{i\varphi}\\
	-\sin\theta e^{-i\varphi}&	\cos\theta e^{-i\gamma}\\
\end{matrix} \right),
\end{equation}
i.e., $(|\Psi'_{1}\rangle,|\Psi'_{2}\rangle)^T=U(|\Psi_{1}\rangle,|\Psi_{2}\rangle)^T$. For the same $P$, all red points are distributed over the blue line. It satisfies $C(\varrho)\leq C(|\Psi_{1}\rangle)+C(|\Psi_{2}\rangle)$. Moreover, the inequality still holds when a random unitary transformation is performed on $|\psi_{1}\rangle$ and $|\psi_{2}\rangle$ simultaneously. Notice that there is an upper limit existing in $C(|\Psi_{1}\rangle)+C(|\Psi_{2}\rangle)$ for each $P$. The upper limit is called the concurrence of assistance (COA) \cite{coa}:
\begin{equation}\label{f23}
C^{a}(\varrho)= \max_{\{p_{i},|\psi_{i}\rangle\}}\sum_{i} p_{i}C(|\psi_{i}\rangle).
\end{equation}

In Fig. \ref{fig2}, red points represent $\big{|}C(|\Psi_{1}\rangle)-C(|\Psi_{2}\rangle)\big{|}$ with  each point indicating one kind of pure state decomposition for $\varrho$. All red points are distributed below the blue line with the same $P$. The inequality $\big{|}C(|\Psi_{1}\rangle)-C(|\Psi_{2}\rangle)\big{|}\leq C(\varrho)$ is correct in this system.

According to Fig. \ref{fig1} and Fig. \ref{fig2}, Eq. (\ref{f14}) is workable in this example.

\section{Discussions and conclusion}

In this paper, we have reviewed the definition of some measures in coherence and entanglement. We provide a general triangle inequality in coherence measures and entanglement concurrence. Then we give an example of entanglement concurrence in a $2$-qubit system. This inequality is formally perfect for its similarity to the triangle inequality in geometry.

Both coherence and entanglement are quantum characteristic and they can be measured for the same state. For a bipartite or multipartite state, coherence is defined as an integral property while entanglement describes the relation among its subsystems. The relation between coherence and entanglement attracts much attention to study. Finding common laws between coherence and entanglement may be an available way to discover more intrinsic connections between them.

Concurrence is a widely used entanglement measure built on the convex-roof construction for mixed states. An attractive question is that whether other coherence or entanglement measures built by the convex-roof construction are suitable for this kind of triangle inequality, such as the entanglement of formation (EOF) \cite{cr1,cr2}, the geometric measure of entanglement (GME) \cite{cr3,cr4}, the convex-roof extended negativity (CREN) \cite{cren}, the G-concurrence \cite{gc} and so on. The right hand side of the inequality Eq. (\ref{f13}) must be correct for the convex-roof construction. But it is an open question that whether the left hand side is workable in other measures.

\section*{ACKNOWLEDGMENTS}
This work is funded by the National Natural Science Foundation of China (Grants No. 11504253 and No. 11474211), the Natural Science Foundation of Jiangsu Province of China (Grant No. BK20141190), the open funding program from Key Laboratory of Quantum Information, CAS (Grant No. KQI201605), and the startup funding from Soochow University (Grant No. Q410800215).

\end{document}